\documentclass[preprint,review,12pt]{elsarticle}

\usepackage{graphicx}

\usepackage{amssymb}
\usepackage{amsbsy}
\usepackage[version=3]{mhchem}

\usepackage[nodots]{numcompress}

\journal{Chemical Physics}

\begin{document}

\begin{frontmatter}


\title{Scattering of a proton with the Li$_4$ cluster: non-adiabatic molecular dynamics description based on 
time-dependent density-functional theory.}



\author[1]{A. Castro}
\ead{acastro@bifi.es}
\author[2]{M. Isla}
\author[3]{Jos{\'e} I. Mart{\'\i}nez}
\author[2]{J. A. Alonso}


\cortext[cor1]{Corresponding author}


\address[1]{Institute for Biocomputation and Physics of Complex Systems (BIFI) and Zaragoza Scientific Center for Advanced Modelling (ZCAM), 
  University of Zaragoza, 50018 Zaragoza, Spain}
\address[2]{Departamento de F{\'\i}sica Te{\'o}rica, At{\'o}mica
y {\'O}ptica, Universidad de Valladolid, 47005 Valladolid, Spain}
\address[3]{Departamento de F{\'{\i}}sica Te{\'{o}}rica de la Materia Condensada,
  Universidad Aut{\'{o}}noma de Madrid, ES-28049 Madrid, Spain}

\begin{abstract}
We have employed non-adiabatic molecular dynamics based on
time-dependent density-functional theory to characterize the
scattering behaviour of a proton with the Li$_4$ cluster. This
technique assumes a classical approximation for the nuclei,
effectively coupled to the quantum electronic system. This
time-dependent theoretical framework accounts, by construction, for
possible charge transfer and ionization processes, as well as
electronic excitations, which may play a role in the non-adiabatic
regime.  We have varied the incidence angles in order to analyze the
possible reaction patterns. The initial proton kinetic energy of 10 eV
is sufficiently high to induce non-adiabatic effects.  For all the
incidence angles considered the proton is scattered away, except in
one interesting case in which one of the Lithium atoms captures it,
forming a LiH molecule.  This theoretical formalism proves to be a
powerful, effective and predictive tool for the analysis of
non-adiabatic processes at the nanoscale.
\end{abstract}

\begin{keyword}
time-dependent density-functional theory
\sep
non-adiabatic molecular dynamics
\sep
ion-cluster collisions
\end{keyword}

\end{frontmatter}


\section{Introduction}
\label{intro}

The study of the interaction of charged particles with matter is a
fundamental area in modern physics, since these collisions are
relevant for many fields of Science. Two relevant examples are
radiation damage in biological tissues, and the stopping power of
solids -- essential in the design, for example, of fusion devices. In
addition to the relatively old areas of ion-atom~\cite{eichler2005}
and ion-surface~\cite{nastasi1996} collisions, the more recent
intermediate discipline of ion-cluster collisions~\cite{connerade2004}
has been developed in the last decades.

In general, these processes -- not only ion-cluster collisions, but
all scattering events of couples of nanoscaled objects, whether they
are atoms, clusters or molecules, charged or not -- may trigger
numerous processes, involving many of the degrees of freedom of the
colliding projectiles: transfer of vibrational energy, transfer of
electronic charge, ionization, fragmentation (sometimes referred to as
collision induced dissociation), recombination, fusion betweeen
clusters, electronic excitations, etc.

Perhaps the most important distinction that can be made is between
\emph{adiabatic} and \emph{non adiabatic} processes. The latter
involve electronic excitations, and this fact implies, from a
theoretical point of view, the necessity of a much more sophisticated
method. If we constrain our consideration to ab initio techniques,
adiabatic processes could in principle be studied with \emph{standard}
adiabatic first principles Molecular Dynamics (MD); however, whenever
charge transfer, ionization, or simply electronic excitations play a
role, some form of non-adiabatic treatment must be used.

One such technique is Ehrenfest MD: it consists of assuming a
classical approximation for the nuclei, that are coupled to the
quantum electronic system. The latter is still a many-particle system
whose out of equilibrium first-principles description is very
demanding. It can be studied, however, ab initio and non-adiabatically
with the help of time-dependent density-functional theory
(TDDFT)~\cite{gross1984,marques2006}, that offers a good balance
between computational effort and accuracy. This idea has received,
among others, the names of Ehrenfest-TDDFT (E-TDDFT), non-adiabatic
quantum MD (NA-QMD), or TDDFT-MD.

The first application of the E-TDDFT equations was however in the
realm of solid state physics~\cite{theilhaber1992}, and in fact with
the purpose of performing adiabatic MD in a different manner. Their
use for non-adiabatic processes was pioneered by Saalmann and
Schmidt~\cite{saalmann1996}; a formal derivation of the model by
Gross, Dobson and Petersilka can be found in Ref.~\cite{gross1996}. It
has later been applied to various collision problems: ion-fullerene
collisions~\cite{kunert2001}, atom-sodium cluster
collisions~\cite{saalmann1998}, charge transfer in atom-cluster
collisions~\cite{knospe1999,knospe2000}, the stopping power of protons
or antiprotons in clusters~\cite{quijada2007,quijada2008} or
insulators~\cite{pruneda2007}, the excitation and ionization of
molecules such as ethylene due to proton
collisions~\cite{wang-zhi-ping2010}, or the interaction of protons or
heavier ions with carbon nanostructures or graphitic
sheets~\cite{krasheninnikov2007,miyamoto2008}. It may also be used to
study laser-induced molecular or cluster dynamics in the high-field
(but still not relativistic) regime; some examples are
Refs.~\cite{castro2004,kunert2005,fennel2010,calvayrac1998,calvayrac1999,handt2006,suraud2000}.


In this work, we have focused on the collision of protons with the
lithium tetramer. We investigate the feasibility of using the E-TDDFT
approach for identifiying the various possible reaction channels:
these collisions are nothing else than chemical reactions with various
possibles outcomes, that depend on the initial velocity, impact
parameter, relative orientations, and even the initial vibrational
state. Depending on the nature of the reactants and on their relative
velocity, the reaction may be non-adiabatic, meaning that the
electronic excited states play a role. A fully unconstrained first
principles study of these chemical reactions in real time is far from
possible -- the complete study would require running over all possible
initial velocities, orientations, etc -- but the E-TDDFT model may
provide some information for selected initial
configurations. Moreover, the current experimental advances in
ultra-fast time-resolved observations of chemical reactions in real
time demand parallel theoretical tools.

In Section~\ref{sec:methodology} we briefly recall the theoretical
framework and the computational methodology; Section~\ref{sec:results}
describes the results; finally we summarize in
Section~\ref{sec:conclusions}.

\begin{figure}
\setlength{\unitlength}{0.99\columnwidth}
\centerline{\includegraphics[width=0.8\unitlength]{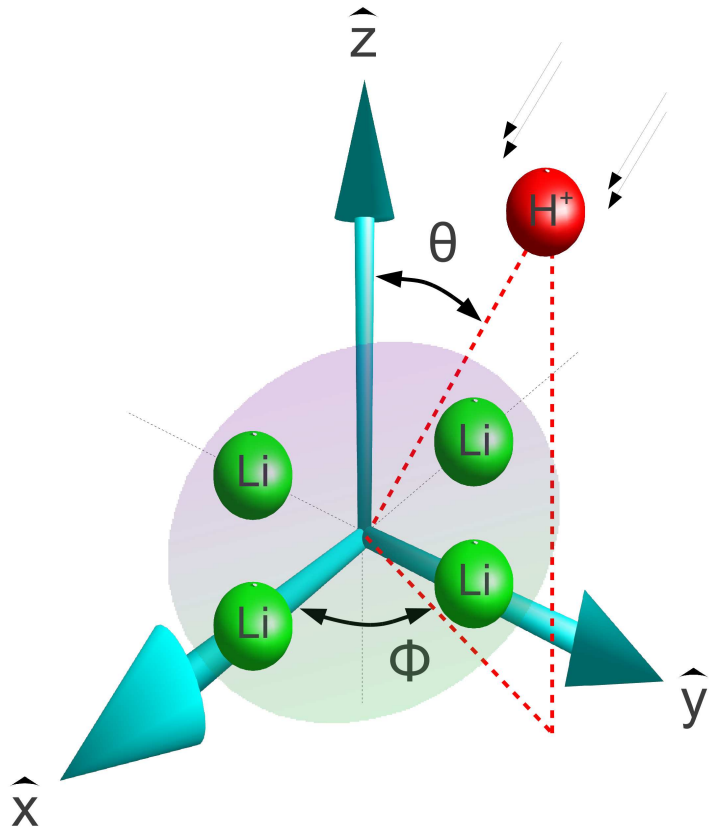}}
\caption{
\label{fig1}
Li$_4$ tetramer (green atoms) and incident proton (red atom). The proton
is always directed towards the center of the tetramer; the incidence angles
are $(\phi,\theta)$.
}
\end{figure}

\section{Methodology}
\label{sec:methodology}

The Ehrefest MD scheme is defined by the following equations (atomic
units are used hereafter):
\begin{eqnarray}
\label{eq:1}
i\frac{\partial}{\partial t}\varphi(x,t) & = & \hat{H}_{\rm e}(R(t))\varphi(x,t)\,,
\\\nonumber
\hspace{-36pt}M_J\frac{{\rm d}^2}{{\rm d}t^2} \vec{R}_J(t) & = & 
-\!\int\!\!\!{\rm d}x\;\varphi^*(x,t)
\nabla_J\hat{H}_{\rm e}(R(t))\varphi(x,t)
\\\label{eq:2}
& & - \nabla_J \sum_{L\ne J}\frac{Z_JZ_L}{\vert\vec{R}_J(t)-\vec{R}_L(t)\vert}\,,
\end{eqnarray}
where $\varphi(x,t)$ is the many-electron wavefunction, that depends
on all the electronic degrees of freedom, denoted $x$. It is governed
by the electronic Hamiltonian $\hat{H}_{\rm e}(R(t))$, which is
determined by all the classical nuclear positions $R(t)\equiv
\lbrace\vec{R}_1(t),\dots,\vec{R}_M(t)\rbrace$.  The motion of the
nuclei is determined by the set of equations~(\ref{eq:2}) -- which are
Newton's equations of motion for each nucleus $J$ (characterized by a
mass $M_J$ and a charge $Z_J$). The electronic Hamiltonian is given
by:
\begin{eqnarray}
\nonumber
\hat{H}_{\rm e}(R(t)) & = & \sum_{j=1}^N\frac{-1}{2}\nabla_j^2 + \sum_{j<k}\frac{1}{\vert\hat{\vec{r}}_j-\hat{\vec{r}}_k\vert}
\\
& & 
- \sum_{Jj}\frac{Z_J}{\vert\vec{R}_J(t)-\hat{\vec{r}}_j\vert}\,.
\end{eqnarray}
The $N$ vector operators $\hat{\vec{r}}_j\; (j=1,\dots,N)$ are the electronic position operators.
This form of the Hamiltonian allows to write the force that acts on
each nucleus solely in terms of the electronic density $n(\vec{r},t)$,
i.e., we can rewrite Eqs.~\ref{eq:2} as:
\begin{eqnarray}
\nonumber
M_J\frac{{\rm d}^2}{{\rm d}t^2} \vec{R}_J(t)  & = & 
-\!\int\!\!\!{\rm d}^3r\;n(\vec{r},t)
\nabla_J v_{\rm ne}(\vec{r},R(t))\,,
\\
& & - \nabla_J \sum_{L\ne J}\frac{Z_JZ_L}{\vert\vec{R}_J(t)-\vec{R}_L(t)\vert}
\end{eqnarray}
\begin{equation}
n(\vec{r},t) = \langle \varphi(t) \vert \sum_{i=1}^N\delta(\vec{r}-\hat{\vec{r}}_j)\vert\varphi(t)\rangle\,.
\end{equation}
The electron-nucleus potential $v_{\rm ne}$ is given by:
\begin{equation}
v_{\rm ne}(\vec{r},R(t)) = 
- \sum_{J}\frac{Z_J}{\vert\vec{R}_J(t)-\vec{r}\vert}\,.
\end{equation}
The possibility of computing the ionic forces solely in terms of the
electronic density permits to use TDDFT, and propagate the proxy
Kohn-Sham (KS) system of non-interacting electrons instead of the real one.
Therefore, we no longer propagate Eq.~\ref{eq:1}, but rather the
time-dependent KS equations:
\begin{equation}
i\frac{\partial}{\partial t}\varphi_i(\vec{r},t) = \lbrace
-\frac{1}{2}\nabla^2 + v_{\rm KS}[n](\vec{r},t)
\rbrace \varphi_i(\vec{r},t)\;\;\;(i=1,\dots,N/2)\,
\end{equation}
\begin{equation}
n(\vec{r},t) = \sum_{i=1}^{N/2} 2\vert\varphi_i(\vec{r},t)\vert^2\,.
\end{equation}
Here, we have assumed an even number of electrons $N$, and a
spin-restricted configuration in which all the KS spatial orbitals
$\varphi_i$ are doubly occupied. The non-interacting electrons move in
the KS potential $v_{\rm KS}[n](\vec{r},t)$, which is divided into the
following terms:
\begin{equation}
v_{\rm KS}[n](\vec{r},t)  = v_{\rm ne}(\vec{r},R(t)) 
+ \int\!\!{\rm d}^3r'\frac{n(\vec{r}',t)}{\vert\vec{r}-\vec{r}'\vert}
+ v_{\rm xc}[n](\vec{r},t)\,.
\end{equation}
The last term, $v_{\rm xc}[n](\vec{r},t)$, is the exchange and
correlation potential, whose precise form is unknown and must be
approximated. In this work we have chosen to use the simplest form,
the adiabatic local-density approximation~\cite{perdew1981}.

We have used the octopus code to run these equations. The numerical
details can be found in Refs~\cite{octopus1,octopus2}; here we will
only summarize the essential aspects: In this TDDFT implementation,
the wave functions, densities and potentials are discretized in a real
space grid, instead of being expanded in basis sets. One important
simplification arises from the use of pseudo-potentials (of the
Troullier-Martins~\cite{troullier1991} type in this work), which
substitute the nucleus and the core electrons of the system by a
smooth set of local and non-local effective potentials, which are seen
by the valence electrons. The Coulomb discontinuity disappears, and
the valence orbitals no longer need to be orthonormal to the core
ones. These facts allow for the use of a real space grid with a
relatively large grid spacing (0.25\AA\; in this work). The molecule
must then be placed in a simulation box, which is a sphere of 20\AA\;
for the simulations that we present here.

The first step in the simulation is the obtention of the ground-state
of the Li$_4$ cluster -- both the lowest energy ionic geometry and the
corresponding electronic structure. The former is a rhombic planar
D$_{2h}$ geometry (see Fig.~\ref{fig1}, the bond length is 2.92\AA). The proton is then placed
18\AA\; away from the centre of the cluster, and is given an initial
relative velocity of 0.02~a.u, which corresponds with a kinetic energy
of 10~eV. The E-TDDFT equations are then propagated, and we have
chosen a total simulation time $T$ of 300~$\hbar$/eV $\approx$ 200~fs.  In
order to numerically propagate the equations, we use the velocity
Verlet algorithm for time-stepping Newton's equations of motion, and
an exponential midpoint rule for the TDKS
equations~\cite{castro2004b}. Part of the electrons may leave the
simulation box during the simulation -- either because of ionization,
or because some of the nuclei may also leave the simulation box,
carrying away some electron density --, and this is accounted for by
making use of absorbing boundaries.

\section{Results}
\label{sec:results}

\begin{figure}
\centerline{\includegraphics[width=0.50\columnwidth]{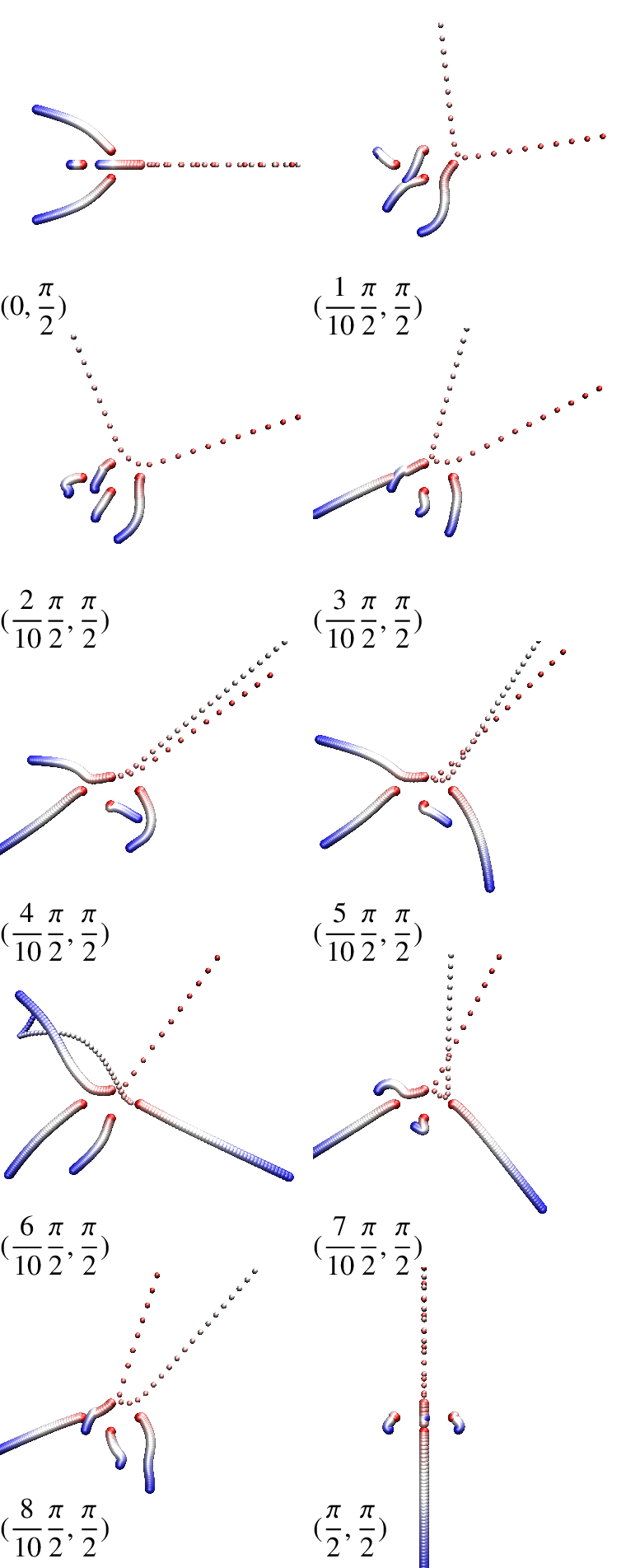}}
\caption{
\label{fig2}
Scattering trajectories for various $(\phi,\theta)$ incidence angles.
Atoms turn from red to white, and then to blue as time progresses.
In this series, the $\theta$ angle is kept fixed at $\pi/2$.
}
\end{figure}

\begin{figure}
\centerline{\includegraphics[width=0.50\columnwidth]{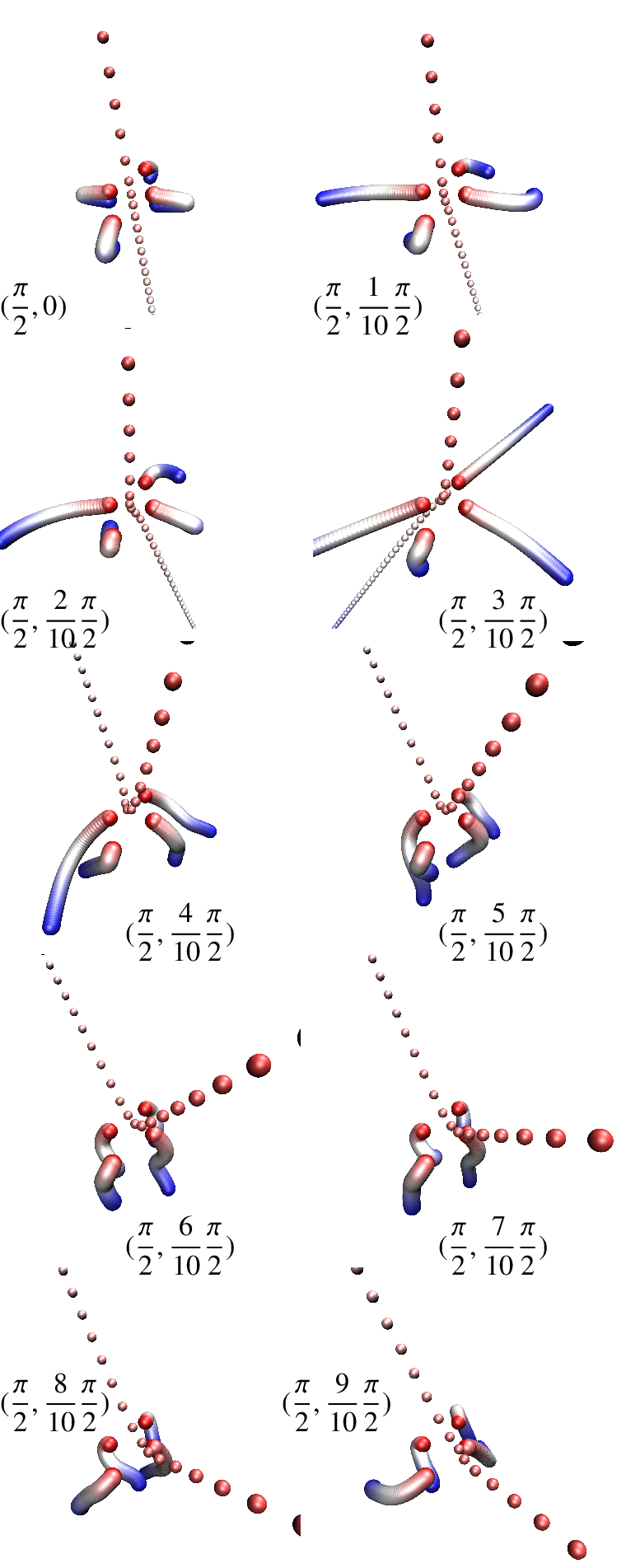}}
\caption{
\label{fig3}
Scattering trajectories for various $(\phi,\theta)$ incidence angles.
Atoms turn from red to white, and then to blue as time progresses.
In this series, the $\phi$ angle is kept fixed at $\pi/2$.
}
\end{figure}

An exhaustive study of this kind of collisions would require the
systematic variation of the following initial conditions: the relative
initial velocity of the colliding fragments, the vibrational state of
the cluster, the incidence angles, and the impact
parameter. Obviously, such a comprehensive study is out of the scope
of any first principles methodology, and one must concentrate on a
given parameter set and range.  In this work, we concentrate on the
incidence angles, since these are perhaps the variables whose small
variations most easily produce different reaction outcomes.

We have simulated the collision of the proton with the lithium
tetramer at a relative velocity of 0.02~a.u, which corresponds with a
kinetic energy of 10~eV. This velocity was selected because it is
sufficiently high to induce non-adiabatic effects (and indeed, we
observed a small but non-negligible amount of ionization in most
cases), but also not too high to induce too simple reactions: at
higher velocities, we mainly observed ``cluster transparency'' (the
proton passes through the cluster loosing part of its kinetic energy,
but otherwise not altering its trajectory), followed in most cases by
Coulomb explosion.

The proton is directed towards the center of the cluster (the
\emph{impact parameter} is therefore zero), and we have varied the
incidence angles $(\phi,\theta)$ (see Fig.~\ref{fig1}), in order to
study how the change in the incidence angle modifies the possible
reaction outcome. We have grouped the simulations in two groups: in
the first group, we have studied in-plane collisions, fixing $\theta$
at $\pi/2$ and varying the $\phi$ angle. These are shown in
Fig.~\ref{fig2}. In the second group, shown in Fig.~\ref{fig3}, the
$\phi$ angle is fixed at $\phi=\pi/2$, and we varied $\theta$.  Each
collision event is represented on a single picture; the complete
trajectory is shown by superimposing all the snapshots (taken at
intervals of $\Delta t \approx 5~$fs), varying the color and tone of
the atoms at different times to make the evolution evident.

In the first series, it can be seen how in all cases except one, the
proton is scattered away at varying angles. For one case, however
($\phi = \frac{6}{10}\frac{\pi}{2}$), one of the Lithium atoms
captures it, and a LiH molecule is formed. Interestingly, the Lithium
that captures the proton is the first one to collide with the proton,
but the capture happens only after the proton is scattered away also
from a second proton. This demonstrates an important fact: in order
for the proton to be captured by the cluster or by any of the
resulting fragments, its energy must be low.  A large incidence
energy, however, deos not rule out the possibility of a capture,
because the proton may loose energy if it is successively scattered by
more than one atom. Other than the LiH, two Li atoms associate and
form a dimer.  In other trajectories of the first series, the Lithium
atoms reorganize in different manners: for example, it is interesting
to see how the $\phi=0$ collision results in Li$_2$ plus two free
Lithium atoms, whereas the apparently similar $\phi=\pi/2$ case
results in Li$_3$ plus only one Lithium atom. In the rest of the
trajectories, the final configuration of positions (and velocities,
which are not displayed) usually permits to predict what would be the
final fragments. However, some cases are perhaps dubious, and longer
simulation times would be needed.

In the second series (Fig.~\ref{fig3}), the proton is scattered away
in all cases. The reason is that the trajectories followed by the
proton pass further away from any of the Lithium nuclei than in some
of the in-plane cases. The first trajectory ($\theta = 0$) is an
example of cluster transparency: the trajectory of the proton is
unaltered, and the Lithium tetramer, although highly excited both
vibrationally and electronically, remains intact. For $\theta > 0$,
i.e., for non-orthogonal collisions, the proton leaves the cluster at
a different angle. For $\theta \ge \frac{5}{10}\frac{\pi}{2}$, the
Li$_4$ also retains its integrity, although strongly distorted. The
other cases display different possible reaction channels.

\begin{figure}
\centerline{\includegraphics[width=0.99\columnwidth]{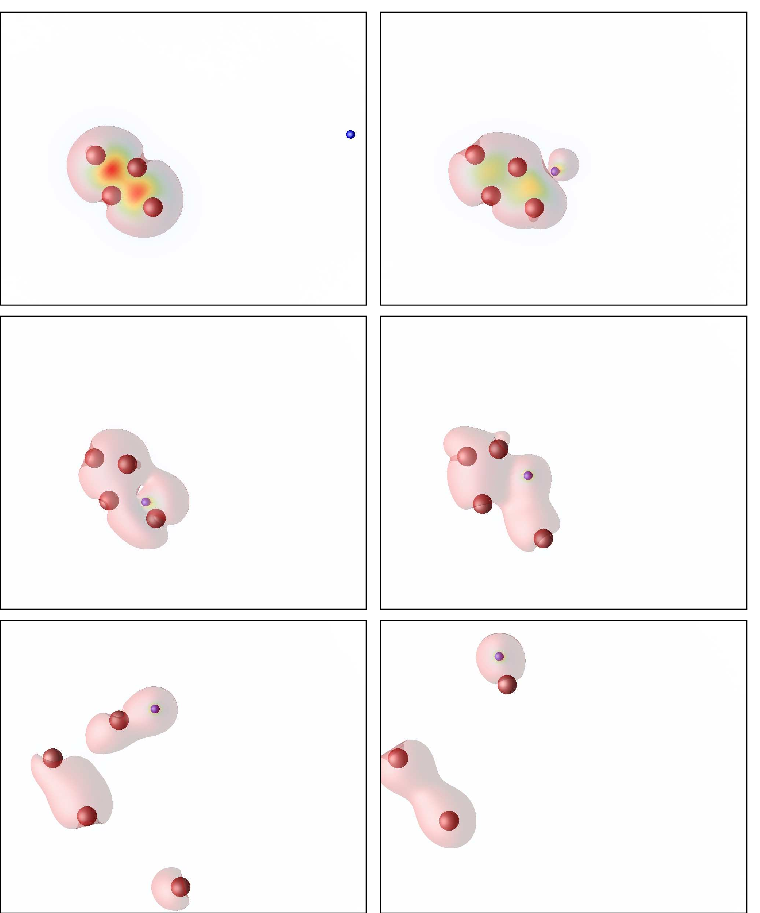}}
\caption{
\label{fig4}
Six snapshot during the scattering trajectories for the $(\phi,\theta) = (\frac{6}{10}\frac{\pi}{2},\frac{\pi}{2})$ incidence angles.
The snapshots correspond with the times $t=0$ (top left), $t=\frac{1}{6}T$ (top right), $t=\frac{2}{6}T$ (middle left), 
$t=\frac{3}{6}T$ (middle right), $t=\frac{4}{6}T$ (botton left) and $t=\frac{5}{6}T$ (botton right). $T\approx 200~$fs.
}
\end{figure}

In all these plots we have not shown the electronic cloud. As an
example, we display in Fig.~\ref{fig4} six snapshots during the
collision at $(\frac{6}{10}\frac{\pi}{2},\frac{\pi}{2})$ incidence
angles. The analysis of the evolution of the charge may help to
predict, at earlier times, whether or not two atoms are to remain
associated: for example, the last snapshot (botton right) clearly
shows the presence of charge between the two Lithium atoms on the left
of the picture, and between the Lithium and the proton on top. This
fact allows to infer with some confidence the formation of stable
bonds (another possibility that would help in this task would be the
use of the time-dependent electron localization function, which is
also accessible with this methodology, see for example
Ref.~\cite{castro2006}).

\section{Conclusions}
\label{sec:conclusions}

We have simulated the scattering of a proton with a lithium tetramer
at moderate energies (10~eV), by making use of non-adibatic molecular
dynamics based on TDDFT. An exhaustive analysis of this process would
require a systematic variation of a large amount of initial
conditions, such as the initial velocity of the colliding fragments,
vibrational state of the target cluster or the impact parameter.  This
is unrealistic for any ab initio approach, but a careful selection of
relevant parameters may provide interesting information: We have
studied, for example, the different reaction channels that occur when
the incidence angles of the proton change. Although long simulation
times are needed in order to identify the final reaction products,
these times are within the computational limits of the first
principles approach that we have followed. In particular, we observed
that for all the incidence angles considered the proton is scattered
away, except in one interesting case in which one of the Lithium atoms
captures it, forming a LiH molecule. In any case, the resulting
collisions reveal interesting information about the scattered proton
and the different reorganization of the Li fragments after the
collisions, providing for each case self-explicative time-evolved
trajectories.  We conclude, therefore, that it is a suitable
methodology to study the various reaction outcomes that result of
proton-cluster collisions, or for any other non-adiabatic collisions
in general.


\section*{Acknowledgements}

This work was supported by MICINN (Spaint) through the grants
FIS2009/13364/C02/01 and MAT2008/06483/C03/01. J. I. M. acknowledges
funding from Spanish MICIIN through Juan de la Cierva Program and
contract FIS2010-16046.






\begin{thebibliography}{00}

\bibitem{eichler2005} J{\"{o}}rg Eichler, ``Lectures on Ion-Atom Collisions'', (Elsevier, Amsterdam, 2005).
\bibitem{nastasi1996} M. Nastasi, J. Mayer, and J. K. Hirvonen, ``Ion-Solid Interactions: Fundamentals and Applications'' 
  (Cambridge University Press, Cambridge, 1996).
\bibitem{connerade2004} ``Latest Advances in Atomic Cluster Collisions'', edited by 
  Jean-Patrick Connerade and Andrey Solov'yov (Imperial College Press, London, 2004).
\bibitem{gross1984} E. Runge and E. K. U. Gross, Phys. Rev. Lett. {\bf 52}, 997 (1984).
\bibitem{marques2006} ``Time Dependent Density Functional Theory'', edited by M. A. L. Marques, 
  C. A. Ullrich, F. Nogueira, A. Rubio, K. Burke, and E. K. U. Gross
(Springer Verlag, Berlin, 2006).
\bibitem{theilhaber1992} J. Theilhaber, Phys. Rev. B {\bf 46}, 12990 (1992).
\bibitem{saalmann1996} U. Saalmann, and R. Schmidt, Z. Phys. D {\bf 38}, 153 (1996).
\bibitem{gross1996} E. K. U. Gross, J. F. Dobson, and M. Petersilka,
  in ``Density Functional Theory'' (Topics in Current Chemistry {\bf
    181}), edited by R. F. Nalewajski, p. 81 (Springer-Verlag,
  Berlin-Heidelberg, 1996).
\bibitem{kunert2001} T. Kunert, and R. Schmidt, Phys. Rev. Lett. {\bf 86}, 5258 (2001).
\bibitem{saalmann1998} U. Saalmann, and R. Schmidt, Phys. Rev. Lett. {\bf 80}, 3213 (1998)
\bibitem{knospe1999} O. Knospe, J. Jellinek, U. Saalmann, and R. Schmidt, Eur. Phys. J. D {\bf 5}, 1 (1999).
\bibitem{knospe2000} O. Knospe, J. Jellinek, U. Saalmann, and R. Schmidt, Phys. Rev. A {\bf 61}, 022715 (2000).
\bibitem{quijada2007} M. Quijada, A. G. Borisov, I. Nagy, R. D{\'{\i}}ez Mui{\~{n}}o, and P. M. Echenique,
Phys. Rev. A {\bf 75}, 042902 (2007).
\bibitem{quijada2008} M. Quijada, A. G. Borisov, and R. D\'\i{}ez Mui\~no, Phys. Stat. Sol. (a) \textbf{205}, 1312 (2008).
\bibitem{pruneda2007} J. M. Pruneda, D. S\'anchez-Portal, A. Arnau, J. I. Juaristi, and E. Artacho, Phys. Rev. Lett. \textbf{99}, 235501 (2007).
\bibitem{wang-zhi-ping2010} W. Zhi-Ping, W. Jing, Z. Feng-Shou,
  Chin. Phys. Lett. {\bf 27}, 053401 (2010).
\bibitem{krasheninnikov2007} A. V. Krasheninnikov, Y. Miyamoto, and D. Tom{\'{a}}nek, Phys. Rev. Lett. {\bf 99}, 016104 (2007).
\bibitem{miyamoto2008} Y. Miyamoto and H. Zhang, Phys. Rev. B {\bf 77}, 161402(R) (2008).
\bibitem{castro2004} A. Castro, M. A. L. Marques, J. A. Alonso, G. F. Bertsch, and Angel Rubio,
  Eur. Phys. J. D {\bf 28}, 211 (2004).
\bibitem{kunert2005} T. Kunert, F. Grossmann, and R. Schmidt, Phys. Rev. A {\bf 72}, 023422 (2005).
\bibitem{fennel2010} T. Fennel, K.-H. Meiwes-Broer, and J. Tiggesb{\"{a}}umker, Rev. Mod. Phys. {\bf 82}, 1793 (2010).
\bibitem{calvayrac1998} F. Calvayrac, P.-G. Reinhard, and E. Suraud, J. Phys. B: At. Mol. Opt. Phys. {\bf 31}, 5023 (1998).
\bibitem{calvayrac1999} F. Calvayrac, P.-G. Reinhard, and E. Suraud, Eur. Phys. J. D {\bf 9}, 389 (1999).
\bibitem{handt2006} J. Handt, T. Kunert, and R. Schmidt, Chem. Phys. Lett. {\bf 428}, 220 (2006).
\bibitem{suraud2000} E. Suraud, and P.-G- Reinhard, Phys. Rev. Lett. {\bf 85}, 2296 (2000).
\bibitem{perdew1981} J. P. Perdew, and A. Zunger, Phys. Rev. B {\bf 23}, 5048 (1981).
\bibitem{octopus1} M. A. L. Marques, A. Castro, G. F. Bertsch, and A. Rubio, Comput. Phys. Comm. \textbf{151}, 60 (2003). 
  See also http://www.tddft.org/programs/octopus.
\bibitem{octopus2} A.~Castro, H.~Appel, M.~Oliveira, C.~A.~Rozzi, X.~Andrade, F.~Lorenzen, E.~K.~U. Gross, M.~A.~L.~Marques and A.~Rubio, {\it Phys. Stat. Solidi B} {\bf 243}, 2465 (2006).
\bibitem{troullier1991} N. Troullier and J. L. Martins, Phys. Rev. B \textbf{43}, 1993 (1991).
\bibitem{castro2004b} A. Castro, M. A. L. Marques, and A. Rubio, J. Chem. Phys. {\bf 121}, 3425 (2004).
\bibitem{castro2006} A. Castro, T. Burnus, M. A. L. Marques, and E. K. U. Gross, in ``Analysis and Control of Ultrafast
  Photoinduced Reactions'', edited by O. K{\"{u}}hn and L. W{\"{o}}ste, Chapter 6.5, pp. 555-576 (Springer-Verlag, Heidelberg, 2007).








\end{thebibliography}



\end{document}